\input psfig.sty

\documentclass{article} 

\begin{document}

\title{SNR0540-69 -the Crab twin- revisited: Chandra vs. HST}

\author{P.A. Caraveo$^1$ \and R.P. Mignani $^2$ \and A. De Luca$^1$ \and S.Wagner$^3$ \and G.F. Bignami $^4$}

\date{ }

\maketitle

\noindent
$^1$ IFC-CNR,  Via Bassini  15,   I-20133 Milan

\noindent
$^2$  ST-ECF, Karl Schwarzschild  Str.2, D8574O   Garching   b.  Munchen

\noindent
$^3$  LaundesSternwarte, Konigstuhl, 69171 Heidelberg

\noindent
$^4$  ASI, Via Liegi 26, I-00148 Rome

\vspace{10mm}

\begin{abstract}
\noindent

High resolution images of the Crab-like SNR SN0540-69 have been recently obtained by the HRC, on board the 
Chandra X-ray Observatory, and by the FORS1 instrument,  attached to ESO/VLT1. A detailed comparison of these 
images with an archival Planetary Camera one will be presented. 
\end{abstract}

\section{Introduction}
The study of extended objects relies heavily on multiwavelength imaging. Irrespective of the astrophysical 
classification of the actual target, its multiwavelength phenomenology provide  precious insights on the physical 
processes at work as well as on the source geometry. However, comparing images taken at different wavelengths is not 
a trivial task . The job is made difficult by the instruments different performances, in particular by their different 
resolution. Although it will always be possible to superimpose a low resolution image on a high resolution  one, the 
most astrophysically compelling comparisons will come from instruments with similar resolutions. The sharpness of 
view of HST, while perfectly suited to map extended sources, sets a very demanding standard for multiwavelength 
comparison. So far, only radio VLA/VLBI images could compete with HST, as shown, e.g. by  Macchetto (1996)\cite{2}.

Unfortunately, the same was not true for X-ray astronomy where the HST performances were, so far, unmatched. The 
High Resolution Camera, on board Chandra Observatory (Weisskopf, O'Dell and van Speybroek 1996\cite{9}),  is the first X 
ray instrument capable of sub arcsec angular resolution, making it possible, for the first time, detailed X-to-optical 
comparisons.

Among the first objects observed by Chandra HRC, during the initial calibration phase, one finds,  apart for the Crab, 
the most Crab-like among the SNRs, SN0540-69, the Crab twin in the LMC.

Since HST did observe the object with the Planetary camera, it is now possible to compare in detail the X and optical 
images of this plerion. Moreover, SN 0540-69 has been observed in polarization mode during the commissioning of the 
FORS1 instrument, now operational at VLT1 (Antu). Owing to the synchrotron emission mechanism generally thought 
to be at work in a plerion, polarization plays a key role in the objects  phenomenology, thus making it desirable to add 
the  polarization information into the multiwavelength study. 0540-69 is a challenging plerion since its LMC location 
implies very reduced angular dimensions,  thus it stands out as a very interesting test case for our multiwavelength 
approach. Here we shall report on the first detailed superposition of  these three data sets: HST, Chandra and 
polarimetry from the ground

\section{Data Analysis}
\subsection{HST} 
SNR B0540-69.3 was observed with the PC chip (45 mas/px) of the WFPC2 on  October  $19^{th}$  1995. Two 300 s 
exposures were obtained through the F555W filter (5442 \AA, 1220 \AA FWHM). The data  have been 
recalibrated by the ST-ECF pipeline using the most recent reference  files and have been combined to reject cosmic ray 
hits.  In order to better study finer details of the remnant structure,  fore/background stars spatially coincident with the 
remnant have been removed by subtracting a model PSF  computed from more than 40 suitable field stars. The 
resulting image is shown in  Fig. 1a, after having being filtered with an adaptive smoothing algorithm with a 4$\sigma$ 
significance threshold. The pulsar (V=22.5) is clearly visible at the center of  the plerion. The plerion morphology 
appears dominated by an elongated structure, on which a secondary emission maximum can be recognized southwest of 
the pulsar.

\subsection{Chandra}
The Chandra X-ray Observatory pointed the supernova remnant SNR B0540-69.3 during the calibration phase of the 
High Resolution Camera (HRC; Murray et al. 1997\cite{6}) on 1999/08/31 for 20 ksec. The data, available in the Chandra 
public Archive (Obs.Id 132), have been used by Gotthelf \& Wang (2000)\cite{1} for a morphological study of the remnant.
This made it possible to resolve three different emission components: (i) a point-like source, identified with the pulsar 
PSRB0540-69, (ii) an elongated  toroidal structure around the pulsar and (iii) a jet like feature apparently protruding 
from the pulsar. As remarked by Gotthelf \& Wang (2000)\cite{1}, when the different distance factor is taken into account, this 
structure is surprisingly similar to the ones observed  by Chandra around the  Crab pulsar.

The HRC data were retrieved from the Chandra Public Archive and reanalyzed using the CIAO software. The image, 
integrated  over the whole HRC energy range (0.08-10 keV) and with the full resolution aspect solution (0.13 
arcsec/px), is shown in Fig.1b, where the three different emission components described above are indicated.

\subsection{VLT}  
Polarimetric observations of SNR 0540-69 were taken on April $12^{th}$ 1999 during the commissioning of FORS1 on the 
VLT/UT1. The instrument was operated in the low-resolution mode, with a corresponding pixel size of 0.2 arcsec. 
Several images with different polarization angles have been obtained under average seeing conditions of 1.2 arcsec and 
combined as described in Wagner \& Seifert (2000)\cite{8}. The processed image, showing the degree of polarization of the 
optical flux in a gray-scale representation, is shown in Fig. 1c. The optical emission from the plerion turns out to be 
significantly polarized, as expected from synchrotron radiation and the polarization maps seem to follow a toroidal 
morphology. Moreover, a clear polarization maximum is also detected around the pulsar position.

\section{The image superposition} 
We have  used the X-ray/optical data to perform a comparative analysis of the remnant morphology. Of course, given 
the small spatial extent of the plerion (5 arcsec), such an analysis must rely on a very accurate image superposition.
Since the optical and X-ray images  are  not directly comparable, a frame registration  procedure as the one described in 
De Luca et al. (these proceedings) can not be applied. We have thus  performed an absolute frame registration using the 
astrometric information attached to the image headers and the ratio between the image pixel sizes as their relative scale 
factor. Unfortunately, such an image superposition yielded residuals of more than 1 arcsec per  coordinate, far too large 
for the study of a 5 arcsec object.  This poor accuracy is obviously traceable to the limited positional accuracy of  the  
reference stars used to define the absolute coordinates of the three data sets.. In  the case of the HST, the absolute 
astrometric accuracy of the Guide  Star Catalog (GSC) ranges typically from 0.2"  to 0.8" per coordinate (Russell  et al. 
1990\cite{7}). Similar values apply for the USNO-A02 catalog, which was used as reference for the VLT observations (Monet 
et al. 1998a,b\cite{4}\cite{5}), while the accuracy of the Chandra HRC aspect solution, depending both on the absolute reference grid 
and on possible errors in the processing of the event files, is expected to be of about 1 arcsec. A better superposition of 
the images can be obtained by registering the coordinates of the common sources in a relative reference frame. To this 
aim it is necessary:
\begin{enumerate}
\item{to select the reference image;}
\item{to compute source centroids with a sufficient accuracy.}
\end{enumerate}
Of course, owing to its excellent angular resolution (pixel size 0.045", FWHM of point sources of  order 0.1"), the HST 
image was obviously chosen as reference and the centroids of possible reference stars were computed with an accuracy 
of 2-4 mas.

\subsection{Chandra vs. HST}
The only source present in the Chandra image is the plerion itself. However, although unresolved  in the Chandra 
image (HRC pixel size 0.132", FWHM 0.5"), the pulsar is bright enough to stand well above the mean count rate from 
the plerion (the peak flux is $3.75\times10^{-3}$ counts pixel$^{-1}$s$^{-1}$,  to be compared with a mean flux of $(0.6\pm0.2)\times 10^{-3}$ counts 
pixel$^{-1}$s$^{-1}$  in an annulus centered on the maximum of emission with inner radius 0.6" and width 1").  This allows the 
determination of the pulsar centroid through 2-D gaussian fitting of its intensity profile, yielding a conservative 
uncertainty of 1 pixel. We then registered the two images by superposing the pulsar centroids in the HST PC reference 
frame. We used as plate scale the ratio between the pixel sizes,  the possible error on this value being not significant 
when propagated over the 2.5" radius of the plerion. To account for image rotation, we used the roll angles of the 
telescopes, whose known accuracies (0.04 deg for HST (FGS instrument handbook) and 0.3 deg for Chandra 
(McDowell 2000\cite{3}) are vastly sufficient not to introduce significant errors, at least in the small region we are studying. 
The result is shown in Fig 2a where we have superimposed theX-ray isophotes to the HST star-free image. 

\begin{figure}
\centerline{\hbox{\psfig{figure=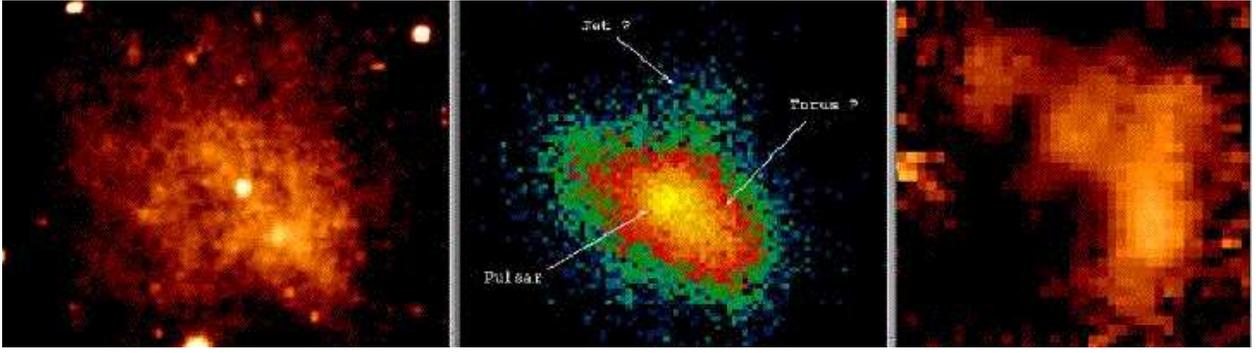,height=4.6cm,clip=}}}  
\caption{a) HST/WFPC2 image of SNR0540-69 obtained through the 555W filter. 
The image is the result of the combination of two 300s exposures. Point 
sources (other than  the pulsar) has been removed by PSF subtraction; b) 
Chandra X-ray observation of the plerion SNR0540-69. The image, integrated 
over the energy range 0.08-10 keV, has been obtained with the High Resolution 
Camera (HRC); c) Polarization map of the plerion. The image, obtained by 
processing several VLT/FORS1 observations in polarimetic mode, shows the 
degree of polarization of the optical flux emitted by SNR0540-69. The pulsar 
is not visible in this map. North to the top, East to the left.}

\end{figure}

\begin{figure}
\centerline{\hbox{\psfig{figure=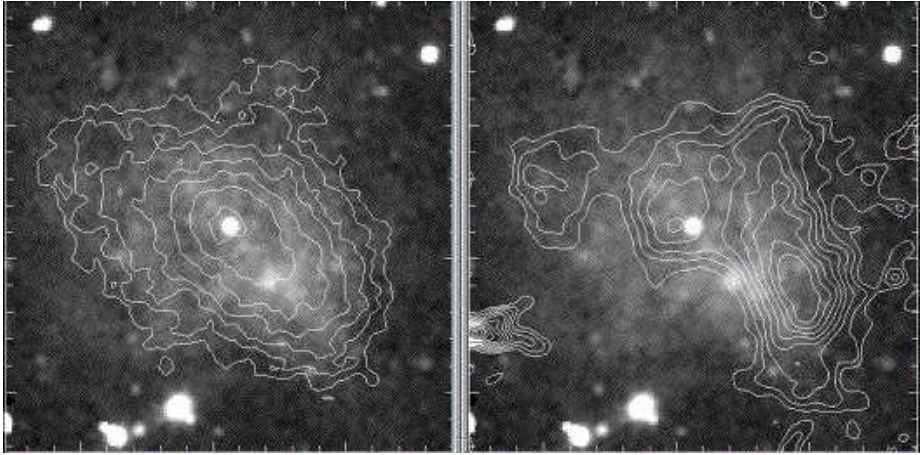,height=6cm,clip=}}}  
\caption{Superposition of the Chandra/HRC(left)  and VLT/FORS1 (right)
contour plots over the HST/WFPC2 image.}
\end{figure}

\subsection{VLT vs. HST}
Since the VLT polarization image contains no resolved point sources, we computed the superposition parameters using 
a total flux image, pertaining to the same set of FORS1 observations in polarimetric mode (see Wagner \& Seifert, 
2000\cite{8}), with the same pointing, orientation and plate scale. In this complementary image we could identify 11 reference 
stars in common with HST image. We then applied the standard astrometric procedure using the centroids of the stars 
to compute the 4 parameters (i.e. X and Y shifts, plate scale and rotation angle) needed to overlay the frames. The mean 
residuals on the reference stars positions turned out to be of order 50 mas, a value mainly driven by the poor seeing of 
the VLT image, however largely sufficient for our purposes. This transformation was then applied to our degree-of-
polarization FORS image and Fig 2b  gives the polarization isophotes superimposed the the HST images.
The registration of the 3 images in a common (the HST PC one) reference frame with an accuracy better than 0.2" 
allows now a careful analysis of the morphological correlations.

\section{The Results}
The HST and Chandra data outline  a very similar elongated pattern featuring a shallow maximum in the SW. Also the 
faint jet, marginally visible in the Chandra image, does have an optical counterpart.  
The polarization image, on the contrary, has a structure of its own, clearly connected with the edges of the elongated 
disk-like plerion, where the maximum of the polarization is detected. A secondary maximum in the polarization 
contours is seen near the pulsar but it does not coincide with it. This is a good example of the importance of accurate 
image superposition in the understanding of extended objects. Proving or disproving correlations at subarcsec level 
between different wavelength images is a challenge that can be met.


\begin{thebibliography}{99}
\bibitem{1}Gotthelf, E.V. \& Wang, Q.D., 2000, ApJ 532, L11
\bibitem{2}Macchetto, D., 1996, in Extragalactic Radio Sources, proc. of 175 IAU Symp., Eckers, R. et al. eds.
\bibitem{3}McDowell, J., 2000, Coordinate Systems for Analysis of On-Orbit Chandra Data, CXC Documentation, 
http://asc.harvard.edu/udocs/docs/docs.html
\bibitem{4}Monet, D. et al., 1998a, Documentation on USNO-A2.0 Catalog, http://www.nofs.navy.mil/projects/pmm/a2.html
\bibitem{5}Monet, D. et al., 1998b, Documentation on USNO-A1.0 catalog, available from ftp.nofs.navy.mil
\bibitem{6}Murray, S.S., et al., 1997, Proc. SPIE, 3114, 11
\bibitem{7}Russell, J.L., et al., 1990, AJ 99, 2059
\bibitem{8}Wagner, S.J. \& Seifert, W., 2000, in Pulsar Astronomy: 2000 and Beyond
\bibitem{9}Weisskopf, M.C., O'Dell, S.L. and van Speybroek, L.P., 1996, Proc. SPIE, 2805, 2


\end{thebibliography}
\end{document}